\documentclass[prl,showpacs, twocolumn,groupedaddress]{revtex4}

\usepackage{graphicx}

\begin{document}

\title
{The Classical-Map Hyper-Netted-Chain (CHNC) method and associated
novel density-functional methods for Warm Dense Matter\footnote{Sanibel Symposium 2011 -novel DFT/WDM}
}

\author{M. W. C. Dharma-wardana}
\email{chandre.dharma-wardana@nrc-cnrc.gc.ca}
\affiliation{
National Research Council of Canada,
Ottawa, Canada K1A 0R6}

\date{\today}
\begin{abstract}
he advent of short-pulse lasers, nanotechnology, as well as
shock-wave techniques have created new states of matter
(e.g., warm dense matter) that call for new theoretical tools.
Ion correlations, electron correlations as well as
bound states, continuum states, partial degeneracies
and quasi-equilibrium systems need
to be addressed. Bogoliubov's ideas of timescales can be used to discuss
the quasi-thermodynamics of non-equilibrium systems.
 A  rigorous approach to the associated many-body problem turns out
to be the computation of the underlying pair-distribution functions
$g_{ee}$, $g_{ei}$ and $g_{ii}$, that directly yield non-local
exchange-correlation potentials, free energies etc., valid within the
timescales of each evolving system.
An accurate classical map of the strongly-quantum uniform
electron-gas problem given by Dharma-wardana and Perrot is reviewed.
This replaces the quantum electrons at $T=0$ by an equivalent
classical fluid at a finite temperature $T_q$, and having the same
correlation energy. It has been shown, but not proven, that the classical
fluid $g_{ij}$ are excellent approximations to the quantum $g_{ij}$.
The classical map
is used with classical molecular
dynamics (CMMD) or hyper-netted-chain integral equations (CHNC)
to determine the pair-distribution
functions (PDFs), and hence their thermodynamic and linear transport properties.
The CHNC is very efficient for calculating the PDFs of
uniform systems, while
CMMD is more adapted to non-uniform systems. Applications to 2D and 3D
quantum fluids,
Si metal-oxide-field-effect transistors, Al plasmas, shock-compressed deuterium,
 two-temperature plasmas, pseudopotentials,
as well as calculations for parabolic quantum dots are reviewed.
\end{abstract}
%
\maketitle
\section{Motivation}
The advent of powerful short-pulse lasers as well as other new tools
for manipulating matter presents  new challenges to existing theory.
Warm dense matter (WDM) is such a regime where we have highly correlated ions,
electrons, finite temperature as well as partial degeneracy effects
that have to be taken into account. Sufficiently thin nano-slabs of
WDM can be studies with a variety of probes~\cite{Ping06,Ng11}.
 Bound states as well as continuum states
have to be treated without sinking in a morass of computations.
The Born-Oppenheimer approximation cannot be used
if coupled-mode effects are important.
 In this paper we examine new theoretical approaches that extend
beyond the  familiar territory of density-functional theory (DFT)
 to treat these and other
intractable problems in many-body physics.

The Hohenberg-Kohn and Mermin (HKM) theorems~\cite{hohen} of
 DFT assert that the
one-body density $n(\vec{r})$ of an inhomogeneous system completely
determines its physics. However,  implementations of
 DFT use the more laborious Kohn-Sham (K-S)
approach~\cite{kohnsham} in lieu of an accurate kinetic-energy
 functional~\cite{PerrotH,Karasiev09}. 
The Kohn-Sham $n(\vec{r})$ of an electron system is:
\begin{equation}
\label{KSdensity.eq}
n(\vec{r})=\sum_\nu |\psi_\nu(\vec{r})|^2f_\nu(\epsilon_\nu/T)
\end{equation}
The K-S eigenstates, $\psi_\nu$ with ``energies'' $\epsilon_\nu$,
occupations factors $f_\nu$ at the temperature $T=1/\beta$ for
all the quantum numbers  $\nu$ have to be determined, self-consistently,
using a {\it one-body} Kohn-Sham potential $V_{KS}$ in the
Kohn-Sham equation. The inclusion of continuum states in this summation
consistently, to satisfy sum rules etc.,
 is a challenge discussed in \cite{dwp82}.
 The Kohn-Sham potential contains an `exchange-correlation potential' $V_{xc}([n])$
that maps the many-body effects to a functional of the one-body
 density. Model $V_{xc}([n])$ potentials have been constructed
 using microscopic theories of systems like the
uniform electron liquid (UEF). Such UEF-calculations
are equivalent to a coupling-constant integration over the
electron-electron  pair distribution function (PDF), viz., $g_{ee}(r)$.
Calculating these PDFS, even for uniform systems, is a challenge that
is treated in this paper.

Quantum systems at high
temperatures behave classically. Then the Kohn-Sham procedure
simplifies. The density  $n(\vec{r})$ is given by the
Boltzmann form:
\begin{equation}
\label{boltz.eq}
n(\vec{r})=n_0\exp\{-\beta V_{KS}(\vec{r})\}
\end{equation}
where $n_0$ is a reference density, and $V_{KS}$ is a classical Kohn-Sham
potential that has to be obtained from a microscopic classical many-body
theory. The `potential of mean-force' used in classical liquid-state
theory is just this classical $V_{KS}$. 
If the center of coordinates is selected to be one of
the classical particles, and if we consider a uniform fluid, then
$n(r)$ becomes the density profile of {\it field particles} around the
 central particle which acts like an external potential. The
density profile  $n(r)$ is
directly related to the pair-distribution
function, i.e.,
\begin{equation} 
n(r)=\overline{n}g(r),\;\;\; \overline{n}=n(r\to\infty).
\end{equation}
Hence one may attempt
to go beyond traditional DFT and proceed directly to the
underlying  calculation of the pair-densities
themselves. The extension of the Hohenberg-Kohn theorem given by
Gilbert, using the one-body reduced density matrix is actually
entirely in this spirit~\cite{Gilbert75}.
 However, the PDF is conceptually easier to
use than the density matrix.
Such considerations suggest that the
kinetic-energy functional may be side-stepped by: 
(i) the use
of an equivalent ``classical-fluid'' at a temperature $T_{cf}$ for
 the (uniform) quantum
fluid whose actual  physical temperature $T$ may even be zero;  (ii)
the use of effective classical  pair-potentials inclusive of quantum effects to
calculate {\it classical} pair distribution functions which can then be
used to compute most of the usual physical properties \cite{prl1}.

The advantage of such a classical-map approach is that the ions, being
essentially classical particles, can be treated together with the electrons
in the same classical computational scheme. Unlike quantum $N$-electron
 schemes which, in principle,
 grow in complexity non-polynomially in $N$,
classical methods are essentially independent of $N$.
Here we should note that traditional
quantum chemistry and condensed-matter physics treat only the electrons 
by DFT. 
On the other hand, Gross and collaborators  have attempted to present a
completely quantum mechanical non-adiabatic DFT theory of electron-nuclear systems,
and given an application to the H$_2$ system \cite{GrossKre01}.
 In standard calculations,
the ion positions are explicitly included and form the external
potential for the motion of the Kohn-Sham electron.
In warm dense matter (WDM), e.g., highly compressed hot hydrogen, there are
as many protons as there are electrons in a given volume of the sample.
Ion motion couples with electron-plasma oscillations to
generate ion-acoustic coupled modes. Their effects may be
missed out in standard DFT formulations as well as in MD simulations.

In any case, the quantum-chemistry approach (e.g., as in the
{\it Gaussian} package) rapidly becomes intractable, esp.
when continuum states have to be included - as in a plasma.
 The solid-state approach of using a
periodic cell is more flexible here, as in the Vienna-simulation package
known as the {\it VASP}.
However, WDM applications demand large unit cells and calculations
of energy bands for many ionic configurations.  The classical-map approach, where both ions
and electrons are treated as classical fluids inclusive of
particle motions,  provides a new paradigm for
warm dense matter and other novel systems which are computationally
very demanding by standard methods. Such standard methods could be
regarded as microscopic bench marks for more global methods like the
classical-map approach discussed here.  

The philosophy of the classical-map technique is to treat the zeroth-order
 Hamiltonian $H_0$ exactly, i.e., using the known quantum solution,
  and then use the classical map
 for dealing with the many-body effects generated from the Coulomb interaction.
  For uniform systems, the eigen-solutions of the
$H_0$ problem are plane waves. Fermi statistics imposes a determinantal form
to the wavefunctions, and hence the non-interacting PDFs $g^0_{ss'}$ are different
from unity if the spin indices $s,s'$ are identical. Thus $g^0_{ss}(r)$ exhibits a
Fermi hole, which can be exactly represented by a classical repulsive potential
known as the Pauli exclusion potential (PEP)~\cite{lado}. This should perhaps be called the
`Fermi-hole potential' as it should not be confused with the `Pauli Potential' defined
in DFT \cite{March86,Trickey09} {\it via} the density-functional derivative of the
difference between the non-interacting kinetic energy and the
 full von Weizs\"{a}cker kinetic energy. In the interest of historical
accuracy, it should however be noted that the name `Pauli potential' was already
in use for the Fermi-hole potential since the work of Lado. We use the names
`Fermi-hole potential' and Pauli-exclusion potential' as synonymous, and different from
the DFT correction to the von Weizs\"{a}cker term known as the Pauli potential.

The key ingredients of the method are the following. 
\begin{enumerate}
\item Replacement of the electron system at
$T$ by a classical Coulomb fluid at an effective classical-fluid temperature given by
\begin{equation} 
\label{tcf.eq}
T_{cf}=(T_q^2+T^2)^{1/2}
\end{equation}
 where $T_q$ is a `quantum temperature'  which depends
only on the electron density. $T_q$  is such that the classical fluid has the
 same correlation energy as the
initial quantum fluid at $T=0$. The motivation for defining $T_{cf}$ by Eq.~\ref{tcf.eq}
is given in \cite{prb2000}.
\item Inclusion of a ``Pauli exclusion potential'', i.e., a Fermi-hole potential
(FHP) to reproduce the Fermi hole of spin-parallel electrons exactly.
\item The 
use of a diffraction-corrected Coulomb interaction $(1/r)(1-e^{-r/\lambda})$
to account for the finite-size of the de Broglie thermal wavelength $\lambda$ of the
electrons at the finite temperature $T_q$.  
\item Calculation of
the pair-distribution functions of the classical fluid using an 
integral-equation method (CHNC), or
 molecular dynamics.  When MD is used in this manner we call it
  classical-map
 molecular dynamics (CMMD). 
\item use of the PDFs in coupling-constant integrations to
calculate the Helmholtz free energy and all other thermodynamic properties of the
 quantum fluid. The linear transport properties (e.g., conductivity) are
  available from Kubo or Ziman-type formulations which use
 the PDFs and potentials as inputs.
\end{enumerate}
Formulations which use this method have been successfully applied
  to a number of quantum systems:\\
(i) The 3-D electron fluid at $T=0$ and at finite $T$ \cite{prl1}. \\
(ii) The 2-D electron fluid both at $T=0$~\cite{prl2,bulutay,prl3,totsuji1}, 
and at finite $T$~\cite{prl3}\\
(iii)
 The calculation of
 Fermi-liquid properties like the electron effective mass $m^*$,  the enhancement of
  the Land\'{e} $g$-factor~\cite{quasi},
 and local-field corrections to the response functions \cite{lfc}.\\  
(iv) The multi-component electron fluid in Si-SiO$_2$ metal-oxide-field-effect
transistors \cite{2valley}; preliminary applications to
multi-valley massless Dirac fermions in graphene \cite{grap07}.\\
(v) Electrons confined in parabolic potentials (quantum dots) \cite{miyake,qdot09}.\\
(vi) Two-mass two-temperature plasmas \cite{cdw-mur}.\\
(vii) Equation of state and Hugoniot of Shock-compress hydrogen \cite{hyd}.\\ 
(viiI) Liquid Al under WDM conditions; linear transport properties of some WDM systems, 
where some of the PDFs were calculated using CHNC \cite{res2006}. 
%
\subsubsection{The QHNC method of Chihara}
\label{qhnc}
For the sake of completeness we also mention Chihara's `quantal-HNC' (QHNC) method~\cite{QHNC}.
 Here an HNC-type equation is solved for the electron subsystem. The electron-electron pair-distribution
 function is calculated by solving the
`quantal HNC equation' with ``a fixed electron'' at the origin. However, the electron
pair-distribution functions obtained by
this method for jellium are in poor agreement with those from
 quantum Monte Carlo methods. In fact, if non-interacting electrons are
considered, the zeroth order PDF, which is known analytically at $T=0$ and
 in terms of a Fermi integral
at finite-$T$ (as discussed below) is not recovered
 correctly by Chihara's method.
The small-$k$ limit of the ion-ion
structure factors calculated by QHNC fail to reproduce the
correct compressibility. Nevertheless, Chihara's QHNC 
recovers some of the short-ranged order in the ion-ion pair-distribution
functions,
where the oscillations and peak heights are in rough agreement
with microscopic simulations. The short-comings in Chira's formulation
 are overcome in the CHNC method.

In the following we discuss details of some of the implementations of CHNC
 using integral-equation methods since they are conceptually more transparent 
and far cheaper than molecular dynamics (CMMD), let alone QMC. 
\section{A Classical representation for the uniform electron liquid}
\label{uef.sec}  
A system of electrons held in place by an external potential (as in a solid,
 a quantum well, or in a molecule) at $T=0$ is necessarily a quantum system.
  The uniform electron
fluid (UEF) at a density $n$, Wigner-Seitz radius $r_s$,
 is the key paradigm for treating exchange and correlation 
in DFT. The pair-distribution functions (PDFs) of the UEF at $T=0$ 
are known from quantum-Monte Carlo (QMC)
studies. They are the basis of exchange-and correlation energies of the UEF.
Hence, if the classical-map scheme could successfully calculate the PDFs of the
electron fluid at arbitrary coupling and spin polarization, in 2-D and 3-D,
then the idea that the quantum fluid can be represented by a classical Coulomb fluid
stands justified. 
 
Consider a fluid of mean density $\overline{n}$ containing
two spin species
with concentrations
 $x_i$ = $\overline{n}_i/\overline{n}$. 
We deal with the physical temperature $T$ of
the UEF, while the temperature $T_{cf}$ of the classical fluid
is $1/\beta$.
Since the leading dependence of the energy on temperature is quadratic,
we construct $T_{cf}$ as in Eq.~\ref{tcf.eq}. This is
clearly valid for $T=0$ and for  high $T$. This assumption has been
examined in greater detail by various applications where it has
been found successful.

The properties of classical fluids interacting via pair potentials $\phi_{ij}(r)$ can be
calculated using classical molecular dynamics (MD) or using an integral equation like the
modified hyper-netted-chain equation. 
The pair-distribution functions for a classical
fluid at an inverse temperature $\beta$ can be written as
\begin{equation}
g_{ij}(r)=exp[-\beta \phi_{ij}(r)
+h_{ij}(r)-c_{ij}(r) + B_{ij}(r)]
\label{hnc}
\end{equation}
Here $\phi_{ij}(r)$ is the pair potential between the
species $i,j$. For two electrons this is
just the Coulomb potential $V_{cou}(r)$.
If the spins are parallel, the Pauli exclusion
principle prevents them from occupying the same spatial orbital.
Following the earlier work, notably by Lado \cite{lado},
 we also introduce a
``Pauli exclusion potential'' or Fermi-hole potential (FHP), ${\cal P}(r)$.
Thus $\phi_{ij}(r)$ becomes ${\cal P}(r)\delta_{ij}+V_{cou}(r)$.
The FHP, ${\cal P}(r)$, is constructed to 
recover the PDFs of the non-interacting UEF, i.e., 
$g^0_{ij}(r)$ is exactly recovered.
The function $h(r)$ = $g(r)$ - 1; it is related to the
structure factor $S(k)$ by a Fourier transform.
The  $c(r)$ is the ``direct correlation function (DCF)''
of the Ornstein-Zernike (OZ)
 equations.
\begin{equation}
\label{oz1}
 h_{ij}(r)  = c_{ij}(r)+
\Sigma_ s\overline{n}_s\int d{\bf r}'h_{i,s}
(|{\bf r}-{\bf r}'|)c_{s,j}({\bf r}')
\end{equation}
The $B_{ij}(r)$ term in  Eq.~\ref{hnc} is
the ``bridge'' term arising from certain cluster interactions.
If this is neglected
Eqs.~\ref{hnc}-\ref{oz1} form a closed set providing the
HNC approximation to the PDF of a classical fluid. 
Since the cluster terms beyond the HNC approximation are
difficult to calculate, they have been modeled approximately
using the theory of hard-sphere liquids \cite{rosen}. 
We  have provided
explicit $B(r)$ functions for the 2-D electron fluid where it
is important even at low coupling~\cite{br2d}.
$B(r)$ is important in 3-D when the
coupling constant $\Gamma=\beta/r_s$ for electron-electron interactions
exceeds, say, 20.  The range of $\Gamma$ relevant
to most WDM work (e.g., $\Gamma \sim 4.5$ even for $r_s$ = 10 )
is such that the HNC-approximation holds well.

\begin{figure}
\vspace{25 pt}
\includegraphics*[width=7.5 cm, height=9.0 cm]{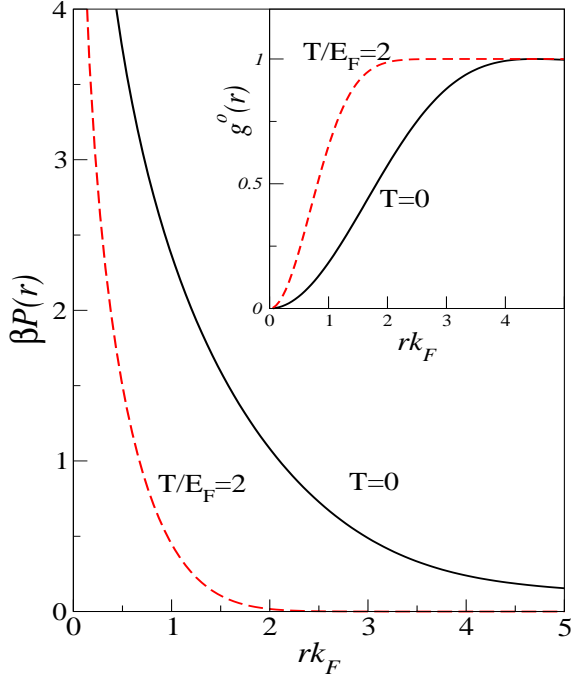}
 \caption
{(Online color) The Ferm-hole potential (i.e., Pauli-exclusion potential)
 $\beta{\cal{P}}(r)$ is a
universal function of $rk_F$ at each $T$ and reproduces the Fermi hole in the parallel-spin
zeroth-order PDF, $g^0_{ss}(r)$, shown in the inset, for $T=0$ and $T/E_F=2$.
If the spins are anti-parallel, $g^0_{s\neq s'}(r)=1$ and the
 Pauli-exclusion potential is zero.}
\label{pau.fig}
\end{figure}

Consider the non-interacting system at temperature $T$, with  $x_i$ = 0.5
for the paramagnetic case.
The parallel-spin PDF,
i.e, $g_{ii}^0(r,T)$, will be denoted by
$g_T^0(r)$ for simplicity, since $g_{ij}^0(r,T)$, i $\ne$ j is unity.
Denoting $({\bf r}_1-{\bf r}_2)$ by ${\bf r}$, it is easy to show, as in
sec.~5.1 of Mahan \cite{Mahan}, that:
\begin{equation}
\label{gzeroT}
g_T^0({\bf r})
=\frac{2}{N^2}\Sigma_{{\bf k}_1,{\bf k}_2}n(k_1)n(k_2)
[1-e^{i({\bf k}_1-{\bf k}_2){\bf \cdot}{\bf r}}].
\end{equation}
Here $n(k)$ is the Fermi occupation number at 
the temperature $T$. Eq.~\ref{gzeroT} reduces to:
\begin{eqnarray}
\label{gzeroTT}
g^0_T(r)&=&1-F_T^2(r)\\
F_T(r)&=&(6\pi^2/k_F^3)\int n(k)\frac{sin(kr)}{r}\frac{kdk}{2\pi^2}.
\end{eqnarray}
Here $k_F$ is the Fermi momentum.
Thus $g^0_T(r)$ is obtained from the Fourier transform of the
Fermi function. The zeroth-order PDF is a universal function
of $rk_F$. It is shown in the inset to Fig.~\ref{pau.fig}.
Assuming that $g^0_{ii}(r)$ can be modeled by an HNC fluid with
the pair interaction $\beta{\cal{P}}(r)$, the ``Fermi-hole
potential'', viz.,  ${\cal{P}}(r)$, 
is easily seen to be given by
\begin{equation}
\label{paudef}
\beta{\cal{P}}(r)=-log[g^0(r)]+h^0(r)-c^0(r)
\end{equation}
The  $c^0(r)$  can be evaluated from
$g^0_T(r)$ using the OZ relations.
The $T=0$ case can be evaluated analytically
\cite{lado}.

We can  determine only the
product $\beta{\cal{P}}(r)$. The classical fluid ``temperature'' $1/\beta$
is still undefined and clearly  {\em cannot} be the thermodynamic
temperature $T$ as $T\to 0$. The Pauli-exclusion potential, i.e., the FHP,
is a universal function of $rk_F$ at each $T$.
It is long ranged and mimics 
the exclusion effects of Fermi statistics that produces
quantum entanglement. 
At finite $T$  the range of the Pauli-exclusion potential 
is comparable to the de Broglie thermal wavelength and
is increasingly hard-sphere like.
Plots of $\beta{\cal{P}}(r)$ and $g^0_{ss}(r)$ are given in 
Fig.~\ref{pau.fig}.

The next step in the CHNC method is to use
 the full pair-potential
$\phi_{ij}(r)$,   
and solve the coupled HNC and OZ equations for the binary 
(up, and down spins) {\it interacting}$\,$ fluid.
For the paramagnetic case,
$\overline{n}_i$ = $\overline{n}/2$, we have:
\begin{eqnarray}
g_{ij}(r)&=&e^{-\beta({\cal{P}}(r)\delta_{ij}+V_{cou}(r))
+h_{ij}(r)-c_{ij}(r)}\\
h_{ij}(q)&=&\stackrel{FT}{\rightarrow}h_{ij}(r)\\
h_{11}(q)&=&c_{11}(q)+(\overline{n}/2)[c_{11}(q)h_{11}(q)+c_{12}(q)h_{21}(q)]\nonumber\\
h_{12}(q)&=&c_{12}(q)+(\overline{n}/2)[c_{11}(q)h_{12}(q)+c_{12}(q)h_{22}(q)]\nonumber\\
   & &     
\end{eqnarray}
The Coulomb potential $V_{cou}(r)$
needs some
discussion. For two point-charge electrons this is  $1/r$.
However, depending
on the temperature $T$, an electron is 
localized to within a thermal wavelength. Thus, following
earlier work, e.g., 
Morita, and Minoo {\it et  al.} \cite{minoo}, we use a
 ``diffraction-corrected'' form:
\begin{equation}
\label{potd}
V_{cou}(r)=(1/r)[1-e^{-r/\lambda_{th}}];\;
\lambda_{th}=(2\pi\overline{m}T_{cf})^{-1/2}.
\end{equation}
Here $\overline{m}$ is the reduced mass of the electron pair, i.e.,
 $m^*(r_s)/2$ a.u., where $m^*(r_s)$ is the electron effective mass.  It
 is weakly $r_s$ dependent, e.g, $\sim$0.96 for $r_s$ = 1. In this
 work we take $m^*$=1. The ``diffraction correction'' ensures the
 correct behaviour of $g_{12}(r\to 0)$ for all $r_s$.

In solving the above equations for a given $r_s$ and at $T$=0,
we have $T_{cf}$=$T_q$. A 
trial $T_q$ is adjusted to obtain an $E_{c}(T_q)$
equal to the known {\it paramagnetic}$\,$ $E_{c}(r_s)$ at each $r_s$, via
a coupling constant integration.
\begin{equation}
E_{xc}(T_q)=\int_0^1\frac{d\lambda}{2}
\int\frac{4\pi r^2dr}{r}[h_{11}(r,\lambda)+h_{12}(r,\lambda)]
\end{equation} 
($E_x$ alone is obtained if $\lambda$ is fixed at 0).
The resulting
 ``quantum''
 temperatures $T_q$  could be fitted to the form:
\begin{equation}
T_q/E_F=1.0/(a+b{\surd{r_s}}+cr_s)
\label{tfit}
\end{equation}
We have also presented a fit to the $T_q$ of the 2-D electron system, and discussed how
the 2-D and 3-D fits could be  related by a dimensional argument.
Bulutay and Tanatar have also examined the CHNC method, and provided fits
to the $T_q$ of the 2-D electron fluid \cite{bulutay}.

\begin{figure}
\vspace{10pt}
\includegraphics*[trim=0.1cm 2.5cm 0.1cm 0.1cm clip-true, width=8.0cm, height=10.5cm]{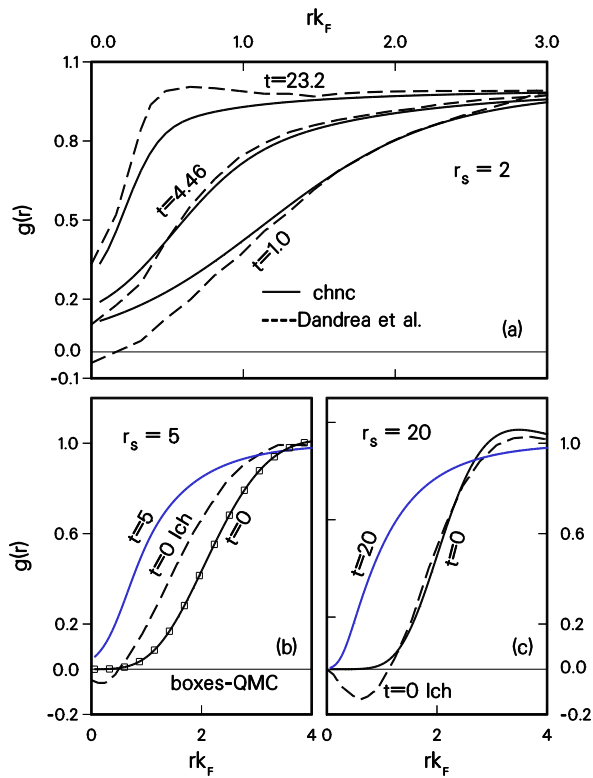}
\caption
{(Online colour) (a) The $g(r)$ from CHNC (solid lines) are compared with
those of Dandrea {\it et al}. \cite{dun}, (dashed lines) at $r_s$=2. The temperature
$t=T/E_F$.
Panel (b) $r_s=5$, CHNC (solid lines) for t=0 snd t=5. The $g(r)$ at $t=0$
from Tanaka and Ichimaru \cite{tan-ichi} (dashed line),
and from DMC-QMC \cite{ortiz94} (boxes), are also shown.
Panel (c) $r_s=5$, CHNC (solid lines) for $t$=0 snd $t$=20. The $g(r)$
of Tanaka and Ichimaru \cite{tan-ichi} (dashed line) is also shown
for $t$=0.
}
\label{grcomp.fig}
\end{figure}

For any given $r_s$, given the $T_q$ from the paramagnetic
case, we can  obtain
$g_{ij}(r)$ and $E_{xc}(r_s,\zeta, T)$ \cite{prl1}, at {\it arbitrary}$\,$
unexplored values of spin-polarization $\zeta$
by solving
the coupled HNC equations, or doing an MD calculation using
the Fermi-hole potential and the diffraction-corrected Coulomb potential.  Many
analytic theories of
electron fluids, e.g., those of Singwi, Tosi, Land and Sj\"{o}lander \cite{Mahan},
Tanaka and Ichimaru, predict $g(r)$ which
become negative for some values of $r$ even for moderate $r_s$.
The PDFs obtained from the HNC-procedure 
are positive definite at all $r_s$. 
In Fig.~\ref{grcomp.fig} we show typical
results for $g_{ij}(r)$ and 
comparisons with QMC-simulations. Our results are
in excellent agreement with the DMC results \cite{prl1}.

The $T_q$ determined from the unpolarized $E_c$ 
is used to calculate $E_{c}(r_s,\zeta, T)$ at any $\zeta$.
The QMC results for $E_{c}(r_s,\zeta)$ at $T=0$
agree with ours, since our $g_{ij}(r)$  agree with those from MC. 
For example, at $r_s$ = 10, the spin-polarized $-E_c$ is:
Ceperley-Alder, 0.0209 Ry; Ortiz-Ballone, 0.0206 Ry \cite{ortiz94};
our method (CHNC), 0.0201 Ry;
Kallio and Piilo, 0.0171 Ry \cite{KP}.

\begin{figure}
\vspace{26 pt}
\includegraphics*[width=7.0 cm, height=7.0 cm]{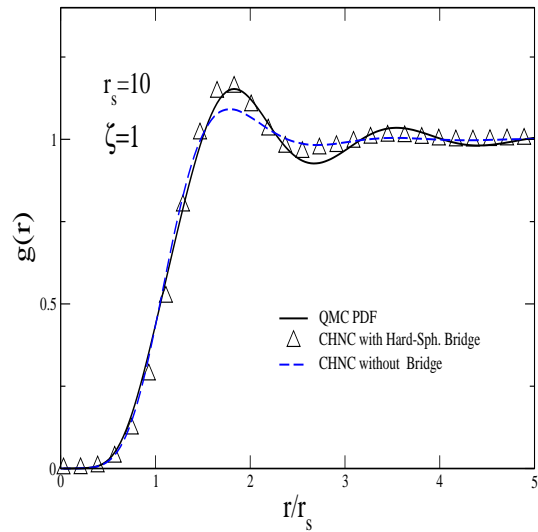}
 \caption
{(Colour online)The QMC pair-distribution function of a fully spin polarized  ($\zeta=1$)
2D electron fluid at $r_s=10$, and $T=0$ is compared with those calculated from CHNC using
a hard-sphere bridge function and with no bridge function what so ever. An essentially exact
fit with QMC can be obtained using a Coulomb bridge function \cite{br2d}.}
\label{gr2d10.fig}
\end{figure}

Most of the recent work using CHNC has been
on the 2D-electron fluid owing to its accrued interest
in nanostructures and technological applications. The electron-electrons interactions
are stronger in reduced dimensions, and the use of a bridge function to supplement the
CHNC equation is essential for accurate work \cite{br2d}. However, even the
appropriately chosen hard-disc bridge works quite well, as seen
in Fig.~\ref{gr2d10.fig}.

\subsection{Fermi-liquid parameters of electron fluids}
\label{fermi-lqd.sub}
It is in fact possible that in some circumstances, WDM may fall into
the category of a Fermi liquid. Highly compressed electron systems have
 correspondingly high Fermi energies
 and hence may have a physical temperature $T << E_F$.
In any case, we review the calculation of Fermi-liquid parameters as
it is an important aspect of the capability of a classical map to
extract results in the strong quantum domain.

Microscopic many-body physics allows one to calculate
various quantities like the effective mass $m^*$ or the Land\'{e} $g$-factor that enter into
Landau's theory of Fermi liquids. One would perhaps assume that a classical representation of
a Fermi liquid would hardly be successful in attacking such problems. For instance, $m^*$ is
usually calculated from the solutions of the Dyson equation for the one-particle interacting
Green's function. If the real part of the retarded self-energy is $\Sigma_1(\vec{k},\omega)$,
the Landau quasi-particle excitation energy $E_{QP}(\vec{k})$, measured with respect to the
chemical potential is used in calculating the effective mass $m^*$.
\begin{eqnarray}
E_{QP}(\vec{k})&=&\epsilon_k + \Sigma_1(\vec{k},\omega)|_{\omega=E_{QP}} \\
\epsilon_k &=& k^2/2-E_F\\
\frac{1}{m^*}&=&\frac{dE_{QP}(k)}{k_F\,dk}|_{k=k_F}
\end{eqnarray}
This is a very arduous calculation, and there are technical questions about the
difficulties of satisfying sum rules, Ward identities etc.,  when the Dyson equation
 is truncated in some approximation.
The values of $m^*$ calculated by different authors using different perturbation
expansions differ significantly, and from QMC results \cite{vigmstr05,quasi}.
\begin{figure}
\includegraphics*[width=9.0cm, height=10.0cm]{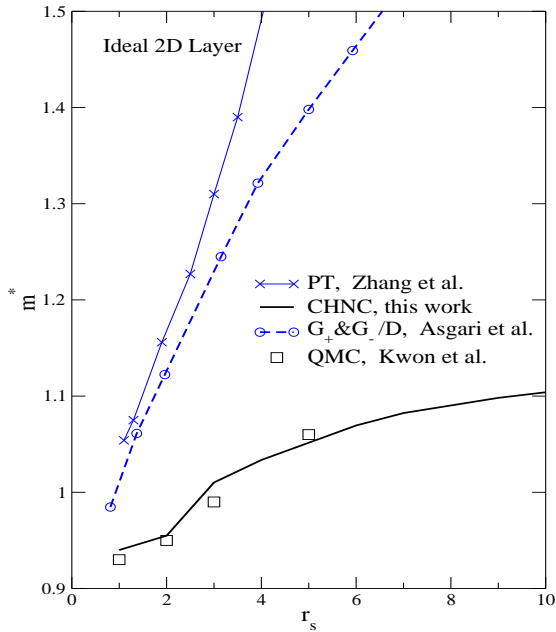}
\caption
{(Colour online) The effective mass $m^*$ of an ideal 2D layer (zero thickness)
 obtained from CHNC are
 compared with the the Quantum Monte Carlo data of Ref.~\cite{kwon} and the
perturbation theory calculations of Zhang {\it et  al.}\cite{zhang},
and Asgari {\it et  al.}\cite{vigmstr05}, i.e., their calculation labeled
$G_+\&G_-/D$. The CHNC calculation is sensitive to the choice
of the bridge function, while the Green's-function methods are
sensitive to the truncations used.
}
\label{m2d-id}
\end{figure}

By contrast, the calculation of $m^*$, and also $g^*$ using CHNC is very simple because it
can evaluate the free energy $F$ of the electron fluid as a function of the physical
temperature $T$ as well as the spin polarization $\zeta$. The ratio of the interacting and
non-interacting specific heats provides a simple evaluation of the $m^*$, while the
ratio of the interacting and non-interacting susceptibilities, determined from the second
derivative (with respect to $\zeta$) of the exchange-correlation correction to the
free energy provides the product $m^*g^*$ \cite{quasi}. 
\begin{eqnarray}
m^*&=&C_v/C_v^0=1+\frac{\left[\partial^2 F_{xc}(t)/\partial t^2\right]}
{\left[\partial^2 F_0(t)/\partial t^2\right]} \\
(m^*g^*)^{-1}&=&\chi_P/\chi_s = 1+
\frac{ \left[\partial^2 F_{xc}(\zeta)/\partial \zeta^2\right]}
 {\left[\partial^2 F_0(\zeta)/\partial \zeta^2\right]}
 \end{eqnarray}
Detailed calculations for ideal 2-D electron layers (see Fig.~\ref{m2d-id}),
 thick layers as well as for
multi-valley systems using the CHNC method have been presented in our
publications \cite{quasi}. Calculations of $m^*$ and $g^*$ for the 3-D electron liquid
using CHNC have not yet been undertaken, while RPA results have been given
by Rice \cite{Mahan}.
\section{Dense Hydrogen and other plasmas}
Dense hydrogen, or any other fully ionized plasma is a direct
 generalization of the uniform electron-fluid problem to include an additional
component (e.g., protons), while removing the positive neutralizing background.
 Let us consider a
fully ionized plasma with ions of charge $\overline{Z}$, and density $\rho$. Then the
electron density $n=\overline{Z}\rho$, and we assume that both subsystems are at the same physical
temperature $T$. 
The electron subsystem will have to be calculated at a classical-fluid temperature
$T_{cf}=\surd(T_q^2+T^2)$ and the electron-electron interactions have to be
diffraction corrected. On the other hand, the ions are classical particles
and the simulations (or integral equations)
for the ions will use the physical temperature $T$. The quantum correction $T_q$
can be neglected for ion, as discussed in \cite{hyd}.
The total Hamiltonian now contains the three terms, $H_i, H_e$, and the electron-ion
interaction $H_{ei}$. The electron system contains two spin components, while the
ion system adds another component. Thus, a three-component problem involving six 
pair-distribution functions have to be calculated. If spin effects could be neglected, then
the two spin-components of the electrons could be replaced by an effective 
one-component electron fluid where the Pauli-exclusion potential (i.e, FHP)
 is included after
averaging over the two components.

\begin{figure}
\includegraphics*[angle=0, width=8.5cm, height=11.0cm]{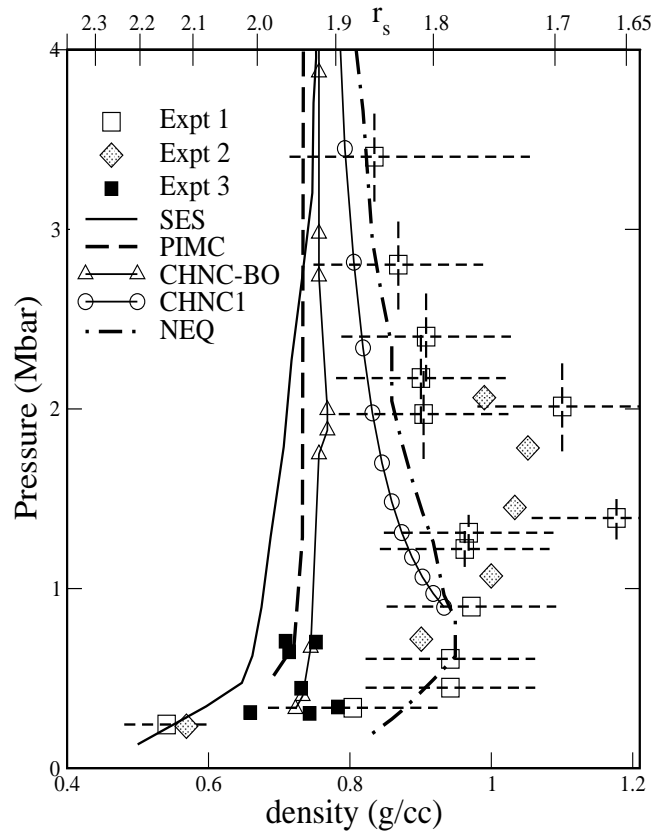}
\caption{
Comparison of the CHNC Hugoniot with experiment and other
  theories for warm-dense Deuterium. Of these, the SESEME (SES),
 path-integral Monte-Carlo (PIMC), 
  CHNC-BO and the CHNC1 are of interest.  A non-equilibrium Hugoniot,
marked NEQ is also shown \cite{hyd}.
  Experiments 1, 2 and 3 refer to Da Silva {\it et al.}, 
Collins  {\it et al.}, and Knudson  {\it et al.}, respectively,
 as described in \cite{hyd}.
}
\label{fighu}
\end{figure}

An example of a classical-map calculation of the EOS of laser-shock compressed hydrogen has been
given by Dharma-wardana and Perrot \cite{hyd}, where a Hugoniot has been calculated
and compared with those from other methods (see Fig.~\ref{fighu}). The article by 
Michael Desjarlais in this issue also refers to the problem of the
 equation of state (EOS) of highly compressed hydrogen \cite{Dejarlais11}.
 A proper experimental probe
of such laser-compression experiments needs to address some method of independent
measurement of the electron temperature $T_e$ and the ion temperature $T_i$.
 If the electrons and ions
are in equilibrium, $T= T_i=T_e$. Then the usual DFT methods using the Born-Oppenheimer
 decoupling would be expected to give a good prediction of the EOS, and  also the Hugoniot. 
The EOS calculation is essentially a calculation of the partition function. This requires the
evaluation of 
\begin{equation}
\label{hei-T.eq}
<e^{-(H_e/T_e + H_i/T_i +H_{ei}/T_{ei})}>
\end{equation}
Here the total Hamiltonian $H$ is rewritten in terms of $H_e$, $H_i$, and the electron-ion
interaction $H_{ei}$ which is again a Coulomb potential. We have included a cross-subsystem
temperature $T_{ei}$ which is simply $T$ for equilibrium systems.  If a Born-Oppenheimer
approximation is used, the electrons `do not know' the temperature of the ions, and
 {\it vice versa}. For equilibrium systems, a Born-Oppenheimer correction can be introduced,
 e.g., as in Morales {\it et  al.} \cite{Morales10}. However, the {\it add on}
correction introduced by Morales et al. will change the virial compressibility,
leaving the small $k\to 0$ behaviour of the proton-proton structure factor
unaffected, and hence the effect on the compressibility sum rules has to be examined. 
In any case
the DFT implementations in codes like VASP, or SIESTA cannot deal correctly with
the case $T_i \ne T_e$, and it is not clear if they treat the $H_{ei}$ term in
 the partition function
correctly even in the equilibrium case, due to the use of the Born-Oppenheimer approximation
which prevents the possibility of coupled electron-ion plasma modes in the system. 

The CHNC technique is a non-dynamical method that
does not need the Born-Oppenheimer approximation. It
correctly treats the cross-interaction $H_{ei}$ even for two-temperature systems,
as established by direct MD simulations \cite{cdw-mur}. Fig.~\ref{fighu} shows that
the SESEME and other standard EOS agree with the CHNC-BO calculation
where $T_{ei}$ is set to $(T_e+T_i)/2$,
while the Laser-shock experiments, where $T_i\ne T_e$ may hold, should agree with 
$T_{ei}$ chosen as  the {\it temperature of the scattering pair}. Ion masses are
much larger than $m_e$, and hence  $T_{ei}$ approaches the electron temperature,
as demonstrated in Dharma-wardana and Murillo  via MD simulations \cite{cdw-mur}.
In effect, the calculation of the Laser-shock hydrogen Hugoniot has to address 
non-equilibrium effects, as well as non-adiabatic effects associated with the
use of the Born-Oppenheimer approximation in standard simulations. 
The conclusions of Galli {\it et  al.} \cite{Galli02} also point to non-equilibrium
effects associated with the electron-ion interaction, i.e., precisely the
term $H_{ei}$ in the Hamiltonian indicated in Eq. \ref{hei-T.eq}. 
Our own views have
evolved beyond what we stated in Ref.~\cite{hyd}, and the subject
probably needs to be revisited, within a two-temperature quasi-equilibrium
setting, without making the Born-Oppenheimer approximation,
especially  at very high compressions.

\subsection{Pseudopotentials}
\label{pseud.sec}
We may also consider the case when the ions are not fully ionized into bare nuclei, but
carry a group of core electrons. For instance, Al-plasmas at 0.5 eV and normal compression
have a charge  $Z=3$ and a core of 10 electrons. Although it is sufficient for
many problems to treat the Al$^{3+}$ as point charges, a more accurate theory may wish to
include the effect of the the core radius and well-depth of the electron-ion interaction
via a pseudopotential. Such pseudopotentials are well known at zero temperature. A very
simple model is that of Ashcroft, while modern implementations are very sophisticated.
 
Al-pseudopotentials suitable for WDM have been given in
parametrized from by Perrot and Dharma-wardana \cite{elr98}. 
The basic idea is to generate the charge density $n(r)$ around a given 
nucleus of charge $Z_N$, immersed in a  UEF of density parameter $r_s$, at
a temperature $T$. The ion is place in a spherical cavity in the positive
background (for details see Ref.~\cite{elr98}) and $n(r)$ is determined by
a Kohn-Sham calculation which satisfies the Friedel sum rule and other
properties. Then we {\it define} a weak non-local pseudopotential $V_{ps}(r)$
by the following relations in $q$-space.
\begin{eqnarray}
\label{pseudLinear.eq}
V_{ps}(q, r_s, T, Z_N)&\equiv&\Delta n(q, r_s, T, Z_N)/\chi(q, r_s,T) \\
\chi(q)&=& \chi(q)^0/\{1-v_q(1-G_q)\chi(q,r_s, T)^0\nonumber
\end{eqnarray}
Here $\chi^0(q,r_s,T)$ is the Lindhard response function at finite $T$ and 
electron Wigner-Seitz radius $r_s$, $v_q=4\pi/q^2$ and $G_q$ is a
local field correction consistent with the density and
temperature of the UEF. Further more, $\Delta n(q, r_s, T, Z_N)$ is the
 Fourier transform of the real-space
free-electron-density pileup $\Delta n(r)$ calculated at the jellium density 
$\overline{n}=3/(4\pi r_s^3)$ and temperature $T$, for the nucleus $Z_n$.
That is
\begin{eqnarray}
\label{den-pileup.eqn}
\Delta n(r)&=& n_f(r)-\overline{n}\\
n_f(r)&=&n(r)-n_b(r) 
\end{eqnarray}
The bound electron density $n_b(r)$ is obtained from the
orbitals of the  finite-$T$ Kohn-Sham
equation as in Perrot \cite{Perrot93}. Here it should be noted that
the bound electrons have to be assigned to a nucleus keeping in mind that some bound states are those
of `hopping electrons' which form a band of localized states near the continuum \cite{hopping92}. 
Equation \ref{pseudLinear.eq} {\it defines}
the pseudopotential to be capable of recovering the charge-pile up via linear response.
Hence it has to render a weak potential.  It is not very satisfactory
if the resulting pseudopotential proves to be
strong. However, the method seems to work in most cases.
 The pseudopotential can usually be parametrized (as in an
Ashcroft empty-core potential), with a core depth $A_0$ and a core radius $r_c$ such that
\begin{eqnarray}
\label{pseud2}
V_{ps}(r)&=&-A_0, \;\;  r<r_c\\
   &=& \overline{Z}/r,  \;\;  r>r_c
\end{eqnarray}
This is evidently a very simple form, compared to modern, hard, non-local
pseudopotentials used in solid-state calculations at $T=0$. 
Such modern potentials remove the core, but a Kohn-Sham equation has
to be solved as they are {\it not weak}, and cannot be treated using linear
 response. However,
we have found that simple potentials as in Eqs.~\ref{pseudLinear.eq}-\ref{pseud2} 
are adequate for even the 
liquid-metal regime close to the melting point, even for metals
which require non-local pseudopotentials at $T=0$. Excellent accuracy is obtained if the
response functions $\chi(q)$ are calculated for
electrons with an effective mass $m^*$ specified for each case. 
It is particularly important to note that the `mean ionization', i.e.,
$\overline{Z}$ is a parameter which appears in the pseudopotential.
The  $\overline{Z}$ is also the Lagrange parameter defining
 the total charge neutrality of the
plasma, as discussed in Refs. \cite{dwp82}, \cite{Perrot93}.
A few examples of this type of simple pseudopotentials are given in
 Table \ref{pseud.tab}.
\begin{table}
\caption{simple pseudopotentials for Al, C, Si at normal
compression, and suitable for the WDM regime, in a.u.}
\begin{ruledtabular}
\begin{tabular}{cccccc}
element & $R_{WS}$ & $\overline{Z}$ & $A_0$ & $r_c$ & $m^*$ \\
\hline\\
Al    &3.141  & 3.0 & 0.3701 & 0.3054    & 0.998 \\
C     &2.718  & 4.0 & 0.0    & 0.3955    & 1.658 \\
Si    &3.073  & 4.0 & 0.0    & 0.9475    & 0.98 \\
\end{tabular}
\end{ruledtabular}
\label{pseud.tab}
\end{table}
The C and Si pseudopotentials were used to generate PDFs of these ionic liquids and
compared with Car-Parinello simulations in Ref.~\cite{csige90}.
Thus these pseudo-potentials can be used in the CHNC equations, or in the CMMD
 simulations, to take account  of the existence  of  a finite-sized core.
Such methods can be used to discuss properties
of warm dense matter, thus providing a complementary approach
to the simulations based on statistical potentials discussed by
Graziani {\it et  al.} in the context of the Cimarron project for
simulations of warm dense matter \cite{Graziani11}.
\section{Two-temperature quasi-equilibria and non-equilibrium systems.}
\label{two-temp.sec}
When energy is deposited rapidly in matter using laser radiation, the electrons absorb the
energy directly and equilibriate among themselves, achieving a very high electron temperature $T_e$.
The ion subsystem, at temperature $T_i$, takes much longer to heat up due to the slow
temperature relaxation via the  electron-ion interaction. Hence, in laser-heated systems, it
is common to find $T_e > T_i$. The inverse situation prevails in shock-heated materials
since the energy of the shock wave couples to the heavy ions and not to the electrons \cite{Ng11}.

The possibility of using a static approach like the CHNC for non-equilibrium
systems resides on Bogoliubov's idea of timescales and conserved quantities.
We have exploited these
ideas  in our work  on hot-electron relaxation, both within Green's-function methods, and within
CHNC methods \cite{elr98}.
The parameters $T_e$, $T_i$ in a two-temperature system are merely Lagrange parameters which assert that,
for certain time scales $\tau_e$, $\tau_i$, the subsystem Hamiltonians $H_e$, $H_i$ are conserved 
quantities. Similarly, a number of other parameters, e.g., quasi-equilibrium chemical potentials,
thermodynamic potentials, pseudopotentials, $\overline{Z}$, etc.,
attached to the subsystems may be conserved for
 the selected time scales. In fact, the original discussions of quasi-equilibria by
Bogoluibov were used in Zubarev's theory of non-equilibrium Green's functions, and RPA-like
results for the quasi-thermodynamics as well as energy relaxation were addressed there-in.
However, RPA-like theories are of limited value. 
In strongly coupled regimes,
the PDFs associated with the given subsystems can be constructed using CHNC, where
the use of the
correct inter-subsystem temperatures (e.g.,  $T_{ei}$)
 for evaluating inter-system PDFS (e.g., $g_{ei}$) is  essential. 
The nature of this inter-system temperature is revealed by its
appearance in the inter-subsystem energy-relaxation formula \cite{elr08}.
A calculation of the distribution functions of two temperature plasmas
 using HNC methods as well
as MD methods was given recently \cite{cdw-mur}.

%
\section{Inhomogeneous systems}
\label{inhom.sec}
The classical-map technique uses a classical fluid at a finite temperature $T_{cf}$ to represent a
uniform-density quantum fluid  at $T=0$. The parameter $T_{cf}$ is density dependent, and hence
the extension to a system with an inhomogeneous density is not straight-forward. Furthermore,
 integral-equation techniques like the HNC become very complicated when applied to inhomogeneous systems.
Molecular dynamics can be applied if a viable mapping can be constructed. However in this connection we
should note that studies of confined classical electrons in
parabolic traps have also yielded useful insights\cite{Wrighton09}.

The classical-map technique treats the zeroth-order Hamiltonian exactly, i.e.,
the map is constructed to reproduce the known quantum solution classically, requiring
the confining potential to be mapped as well. Even when there is no confining 
potential (other than a uniform background), the zeroth-order problem of  $H^0$ has to 
be correctly treated.  This was done in the
UEF problem by constructing a Pauli exclusion potential (i.e., the Fermi-hole potential)
 to recover the Fermi hole in the
$g^0_{ss}(r)$ exactly. 

When non-interacting electrons are placed in an external potential, e.g., a parabolic trap,
the uniform density $\overline{n}$ modifies to a new distribution $n^0(r)$. 
Classically, this distribution
is of the Boltzmann form, Eq. \ref{boltz.eq} where $V_{KS}(r)$ contains all the terms found
in the
exponent of the HNC equation. Thus, given the $n^0(r)$ calculated from a quantum mechanical
treatment of $H_0$ which contains the parabolic external potential, it is necessary to
invert the HNC equation to get the effective classical potential which corresponds to $n^0(r)$.
A simplified approach to this was used by us in ref.~\cite{qdot09}.
 At this stage the calculation is
somewhat similar to the determination of the Pauli exclusion potential, and hence
the specification of an effective fluid temperature does
not become necessary. 
The classical Coulomb fluid at a finite temperature $T_{cf}$ is still necessary
 for dealing with the many-body effects generated from the Coulomb interaction. However, given a
 non-uniform distribution, there is no evident method of defining a unique $T_{cf}$ and the
 simplicity of the original CHNC method is lost. Further more, the electron-electron 
 pair-distribution functions now depend explicitly on two coordinates, viz., 
 $g(\vec{r}_1,\vec{r}_2)_{ss'}$. The use of molecular-dynamics simulations is more
 convenient in dealing with systems where the simplicity of homogeneous systems is
 lost. Another advantage  an MD simulation is that the
 the bridge-function approximations are  avoided.
 
 In mapping an inhomogeneous system of density $n(r)$ to a homogeneous slab of density $\overline{n}$
 we have used the form \cite{quasi,Jost05,qdot09}, 
 \begin{equation}
 \overline{n}=<n(r)n(r)><n(r)>
 \end{equation}
in dealing with 2D systems. The same method has been used
 by Gori-Giorgi and Savin for 3D systems \cite{ggsavin}.
Using such a uniform density to define a unique temperature of an equivalent classical
fluid, we were able to reproduce the charge distribution of interacting electrons in
2D quantum dots obtained from Quantum-Monte Carlo methods. However, as we used CHNC, it was 
necessary to introduce bridge-functions and boundary corrections which impaired the
 transparency of the classical map. 
Hence this work \cite{qdot09}  may be regarded as a preliminary attempt.

\section{Conclusion}
We have outlined the classical-map technique of treating the quantum many-body problem in Fermi
systems via a mapping to an equivalent classical system at a density-dependent effective temperature
different from the physical temperature,
and where the particles interact by a
pair potential consisting of a Pauli-exclusion potential and a diffraction-corrected
Coulomb potential. Large numbers of particles, and their thermodynamics or quasi-thermodynamics can be easily 
calculated. Since pair-distribution functions can be calculated accurately, and at any value of the coupling
constant, the adiabatic connection formula provides results for the non-local exchange-correlation functionals
in an entirely unambiguous, rigorous manner. No gradient corrections, meta-functionals etc., are needed.
The Born-Oppenheimer approximation is not necessary as the CHNC technique is
not dynamical.
Hence the method would be of great interest from the point of view of equations-of-state studies, both for
equilibrium, and for quasi-equilibrium systems.

Since suitable derivatives of the free energy with respect to density, temperature, and spin polarization
lead to Landau Fermi-liquid parameters, the method is capable of easily furnishing alternative
results for the effective mass $m^*$, Land\'{e}-$g$ factor, local-field factors of response functions
etc., which are difficult to determine by standard Greens-function perturbation techniques of quantum
many-body theory.

The application of the method to inhomogeneous systems is still poorly developed. Similarly, the
method, being a technique for the total energy as a functional of the pair density, is
similar to DFT in not yielding spectral information within its own formal structure.


\begin{thebibliography}{99}

\bibitem{Ping06}
Y. Ping, D. Hanson, I. Koslow, T. Ogitsu, D. Prendergast, E. Schwegler, G. Collins and Andew Ng.
 Phys. Rev. Lett. {\bf 96} 255003 (2006)

\bibitem{Ng11}
Andre Ng, \url{http://www.qtp.ufl.edu/sanibel/topics.shtml} Sanibel Symposium. (2011)


\bibitem{hohen}P. Hohenberg and W. Kohn, Phys. Rev. {\bf 136}, B864 (1964);
 D. Mermin, Phys. Rev. {\bf 137}, A1441 (1965)

\bibitem{kohnsham} W. Kohn and L.J. Sham, Phys. Rev. {\bf 140}, A1133 (1965)

\bibitem{PerrotH}F. Perrot, J. Phys.: Condens. matter {\bf 6}, 431 (1994) 

\bibitem{Karasiev09}
V.V. Karasiev, R.S. Jones, S.B. Trickey and Frank E. Harris, {\it New developments in Quantum Chemistry},
Eds. Jos\'{e} Lous Paz, . J. Hern\'{a}ndez, p25-54 (2009)

\bibitem{dwp82}
M.W.C. Dharma-wardana and F. Perrot,  Phys. Rev. A {\bf 26}, 2096 (1982) 

\bibitem{Gilbert75}
T. L. Gilbert, Phys. Rev. B {\bf 12}, 2111 (1975)

\bibitem{GrossKre01}
T. Kreibich and E.K.U. Gross. Phys. Rev. Lett., {\bf 86}, 2984, (2001).

\bibitem{lado}
 F. Lado, J. Chem. Phys. {\bf 47}, 5369 (1967)

\bibitem{March86}
N. H. March, Phys. Lett. A {\bf 113} 476 (1986)

\bibitem{Trickey09}
S. B. Trickey, V. V. Karasiev,  R. S. Jones, Int. J. Q. Chem. {\bf 109}, 2951 (2009)  


\bibitem{prl1}
M. W. C. Dharma-wardana and F. Perrot, Phys. Rev. Lett. {\bf 84}, 959 (2000) 

\bibitem{prb2000}
F. Perrot and M.W.C. Dharma-wardana,  Phys. Rev. B15  {\bf 62}, 16536 (2000)
 {\it Erratum} {\bf 67}, 79901 (2003)                                               

\bibitem{prl2}
Fran\c{c}ois Perrot and M. W. C. Dharma-wardana,  Phys. Rev. Lett. {\bf 87},
 206404 (2001)         

\bibitem{bulutay}
C. Bulutay and B. Tanatar, Phys. Rev. B {\bf 65}, 195116 (2002)

\bibitem{prl3}
M. W. C. Dharma-wardana and F. Perrot, Phys. Rev. Lett. {\bf 90}, 136601 (2003)

\bibitem{totsuji1}N. Q. Khanh and H. Totsuji, Solid State Com., {\bf 129}, 37 (2004)


\bibitem{quasi}
M. W. C. Dharma-wardana, Phys. Rev. B  {\bf 72}, 125339 (2005)

\bibitem{lfc}
M. W. C. Dharma-wardana and F. Perrot., Europhys. lett.  {\bf 63}, 660  (2003)

\bibitem{2valley}
M. W. C. Dharma-wardana and F. Perrot, Phys. Rev. B {\bf 70}, 035308 (2004)

\bibitem{grap07}
M. W. C. Dharma-wardana,  Phys. Rev. B {\bf 75}, 075427 (2007)

\bibitem{miyake}
T. Miyake, C. Totsuji, K. Nakanishi, and H. Totsuji, Phys. Let. A {\bf 372}, 6197 (2008).

\bibitem{qdot09}
M. W. C. Dharma-wardana , Physica E {\bf 41},  1285 (2009)

\bibitem{cdw-mur}M. W. C. Dharma-wardana and M. S. Murillo,
 Phys. Rev. E.  {\bf 77}, 026401 (2008)

\bibitem{hyd}
M. W. C. Dharma-wardana and F. Perrot, Phys. Rev. B, {\bf 66}, 14110 (2002)

\bibitem{res2006}
M. W. C. Dharma-wardana,  Phys. Rev. E {\bf 73},  036401 (2006)

\bibitem{QHNC}
 J. Chihara and S. Kambayashi, J. Phys: Condens. matter {\bf 6} 10221 (1994)


\bibitem{rosen}
Y. Rosenfeld and N.W. Ashcroft, Phys. Rev. A {\bf 20}, 2162 (1979)


\bibitem{br2d}
M. W. C. Dharma-wardana, Phys. Rev. B 82, 195303 (2010)

\bibitem{Mahan}
G. D. Mahan, {\it Many-particle physics}, Plenum Press, New York (1990)

\bibitem{minoo}
M. Minoo, M. Gombert and C. Deutsch, Phys. Rev. A {\bf 23}, 924 (1981)


\bibitem{dun}
R. B. Dandrea, N. W. Ashcroft and A. E. Carlsson, Phy. Rev. B
{\bf 34}, 2097 (1986).

\bibitem{tan-ichi}
S. Tanaka and S. Ichimaru,
Phys. Rev. B {\bf 39}, 1036 (1989)


\bibitem{ortiz94}
G. Ortiz abd P. Ballone, Phys. Rev. B {\bf 50}, 1391 (1994)

\bibitem{KP}
A. Kallio and J. Piilo, Phys. Rev. Lett. {\bf 77}, 4237 (1996)


\bibitem{vigmstr05}
R. Asgari, B. Davoudi, M. Polini, M. P. Tosi, G. F. Giuliani, and
G. Vignale, Phys. Rev. B {\bf 71}, 45323 (2005).
 
\bibitem{kwon}
Y. Kwon, D. M. Ceperley, and R. M. Martin, Phys. Rev. B {\bf 50}, 1684 (1994)

\bibitem{zhang}
Y. Zhang and S. Das Sarma, Phys. Rev. B {\bf 71},45322 (2005)

\bibitem{Dejarlais11}
M. Desjarlais \url{http://www.qtp.ufl.edu/sanibel/topics.shtml} Sanibel Symposium. (2011)

\bibitem{Morales10}
Miguel A. Morales, Carlo Pierleoni, and D. M. Ceperley,  Phys. Rev. B {\bf 81}, 021202 (2010) 

\bibitem{Galli02}
Giulia Galli, Randolph Q. Hood, Andrew U. Hazi, and Fran\c{c}ois Gygi, Phys. Rev. B {\bf 61},
909 (2002); F. Gygi and G. Galli, {\bf 65}. 220102 (2002) 

\bibitem{Perrot93}  F. Perrot, Phys. Rev. A {\bf 47}  570 (1993);
F. Perrot and M.W.C. Dharma-wardana,   Phys. Rev. E. {\bf 52},
  5352 (1995).

\bibitem{hopping92}
M.W.C. Dharma-wardana and F. Perrot, Phys. Rev. A {\bf 45},5883 (1992).

\bibitem{Graziani11}
F. R. Graziani {\it et  al.} Lawrence Livermore National Laboratory report, USA,  LLNL-JRNL-469771 (2011)

\bibitem{csige90}
 M.W.C. Dharma-wardana and F. Perrot,
 Phys. Res. Lett. {\bf 65}, 76 (1990)

\bibitem{elr08}
M. W. C. Dharma-wardana, Phys. Rev. Lett. {\bf 101}, 035002 (2008)

\bibitem{Wrighton09}
J. Wrighton, J. W. Dufty, H. K\"{a}hlert and M. Bonitz, Phys. Rev. E {\bf 80}, 066405 (2009)


\bibitem{elr98}
 M. W. C. Dharma-wardana and F. Perrot, Phys. Rev. E, {\bf 58}, 3705
  (1998);{\it Errratum} Phys. Rev. E {\bf 63}, 069901 (2001)


 \bibitem{Jost05}
 D. Jost and M. W. C. Dharma-wardana,
 Phys. Rev. B, {\bf 72}, 195315 (2005)


\bibitem{ggsavin}
 P. Gori-Giorgi and A. Savin, 
Phys. Rev. A {\bf 71}
 32513 (2005)


\end{thebibliography}
\end{document}